# PROGRAMS AND ALGORITHMS FOR THE SHELL DECOMPOSITION OF OSCILLATING FUNCTIONS IN SPACE

by

**Ludmila Urzhumtseva[§], Vladimir Y. Lunin[&] & Alexandre Urzhumtsev[#€]**

[§]*Architecture et Réactivité de l'ARN, UPR 9002 CNRS, IBMC (Institute of Molecular and Cellular Biology), 15 rue R. Descartes, 67084 Strasbourg, France ; Université de Strasbourg, Strasbourg, F-67000 France*

[&]*Institute of Mathematical Problems of Biology RAS, Keldysh Institute of Applied Mathematics of Russian Academy of Sciences; Pushchino, 142290, Russia.*

[#]*Centre for Integrative Biology, Institut de Génétique et de Biologie Moléculaire et Cellulaire, CNRS–INSERM-UdS, 1 rue Laurent Fries, BP 10142, 67404 Illkirch, France*

[€]*Université de Lorraine, Faculté des Sciences et Technologies, BP 239, 54506 Vandoeuvre-les-Nancy, France*

*E-mail:* l.ourjoumtseva@ibmc-cnrs.unistra.fr , lunin@impb.ru , sacha@igbmc.fr

**Abstract**

Real-space refinement of atomic models in macromolecular crystallography or in cryo electron microscopy fits a model to a map obtained experimentally. This requires generating model maps of a limited resolution which moreover may vary from one molecular region to another. Calculating such map as a sum of atomic contributions requires that these contributions reflect the local resolution of the experimental map. A possibility to refine the parameters of these contribution means to express it as a function of atomic coordinates, displacement factor and eventually of resolution. Recently, Urzhumtsev & Lunin (*BioRxiv*, 10.1101/2022.03.28.486044) suggested to decompose finite-resolution atomic images, and more generally spherically symmetric oscillating functions in space, into a sum of specially designed terms analytically dependent on all atomic parameters. Each term is a spherically symmetric function concentrated in a spherical shell. Here we describe the software and respective algorithms to carry out such shell decomposition of oscillating functions.

**Synopsis**: A software has been developed to decompose spherically symmetric oscillating functions in space into a sum of specially designed shell-shaped functions. Such decomposition allows one to represent a resolution-restricted atomic image by a function dependent analytically on the resolution, the isotropic disorder parameter, and the atomic coordinates.

**Introduction**

In macromolecular crystallography (MX) or in cryo-electron microscopy (cryoEM), real-space refinement of atomic models, *i.e.*, their fit to the experimental or experimentally based maps, is the key step of model improvement and validation (*e.g.*, Diamond, 1971; Chapman *et al*., 2013; Afonine *et al*., 2018; Urzhumtsev & Lunin, 2019). This step becomes even more important (Palmer & Aylett, 2022) with the recent progress in cryoCM (Kühlbrandt, 2014) and in prediction of protein structures (Jumper *et al.,* 2021; Baek *et al*., 2021).

Three-dimensional fields such as an electrostatic potential or an electron or nuclear density can be seen as a sum of contributions from individual independent atoms. The experimental maps of these fields are their distorted images, basically due to an atomic disorder and a limited resolution. To be compared quantitatively with an experimental map, a map obtained from an atomic model should mimic the same distortions. Such map can be calculated as a sum of independent atomic contributions considering respectively both the atomic disorder and the resolution cut-off with which the experimental maps are available (*e.g*., Chapman, 1995; Lutdke *et al*., 1999; Chapman *et al*., 2013; DiMaio *et al*., 2015, Sorzano *et al*., 2015, and references therein). In the simplest case of an isotropic harmonic disorder, such contributions, also called atomic images in the given map, are spherically symmetric oscillating functions. They are composed of a peak in the origin surrounded by so-called Fourier ripples with their amplitude decreasing to zero much slower than an atomic density itself. These functions, in particular the size and position of the Fourier ripples, depend on the atomic type, resolution and the disorder amplitude. Modeling the central peaks only, *e.g.*, Lunin & Urzhumtsev (1984), is insufficient to reproduce the map accurately.

Spherically symmetric oscillating functions in space, in particular atomic images, can be decomposed into a sum of terms, each represented by a specially designed spherically symmetric analytical function ($\Omega$-function) that describes a distribution concentrated in a thin spherical shell (Urzhumtsev & Lunin, 2022). The main feature of the $\Omega$-function is that it does not change its form when changing the displacement parameter or the image resolution but only modifies the value of the respective parameters. This allows an analytic expression of the model map values via all parameters of the atomic model including the local resolution, now associated with the atom (Urzhumtsev & Lunin, 2022). This in turn assures analytic expressions for all respective partial derivatives of the score function comparing the maps and, as result, an efficient model refinement.

In this work, we describe the algorithms and the programs that realize such shell decomposition of the oscillating functions in space. In what follows, we use the term "density" for all scalar fields since the mathematical and computer tools are exactly the same whatever the physical meaning of the function is. The term "atomic density" stands for the density distribution of an isolated atom and "atomic image" describes how this density is seen in a given map, *i.e.*, considering both an atomic disorder and a respective resolution cut-off.

## 1. Algorithms

*1.1. General considerations*

Introducing a harmonic disorder into an atomic position leads to blurring the atomic density function by its convolution with the normalized Gaussian function

$$g(\mathbf{r}; B) = \left(\frac{4\pi}{B}\right)^{3/2} \exp\left(-\frac{4\pi^2 |\mathbf{r}|^2}{B}\right) \qquad (1)$$

The parameter $B = B_0$ is the atomic displacement factor. The density of an immobile atom is a spherically-symmetric function with a peak in the origin and decreasing monotonously and quite sharply with the distance. Traditionally, it is decomposed into a linear combination of Gaussian functions (*e.g.*, Doyle & Turner, 1968)

$$f(\mathbf{r}) \approx \sum_{m=1}^{M} C_m g(\mathbf{r}; B_m) \qquad (2)$$

This allows one to use the property of the Gaussian function

$$g(\mathbf{r}; B) * g(\mathbf{r}; B_0) = g(\mathbf{r}; B + B_0) \qquad (3)$$

to express analytically the atomic density for any $B_0$ value.

To represent oscillating spherically symmetric function $f(\mathbf{r})$ in space we suggested the decomposition

$$f(\mathbf{r}) \approx \sum_{m=1}^{M} C_m \Omega(\mathbf{r}; R_m, B_m) \qquad (4)$$

where

$$\Omega(\mathbf{r}; R, B) = \frac{1}{|\mathbf{r}|R} \sqrt{\frac{1}{4\pi B}} \left[ \exp\left(-\frac{4\pi^2(|\mathbf{r}| - R)^2}{B}\right) - \exp\left(-\frac{4\pi^2(|\mathbf{r}| + R)^2}{B}\right) \right] \qquad (5)$$

Function $\Omega(\mathbf{r}; R, B)$ represents a spherically symmetric solitary wave in three-dimensional space (**Fig. 1**). It can be thought as a density of a virtual unit charge distributed uniformly on the spherical surface of the radius $R$ and blurred by the isotropic harmonic disorder described by the parameter $B$ (Urzhumtsev & Lunin, 2022). Decomposition (4) of atomic images, which we call the shell decomposition, is a generalization of the decomposition (2) because

$$\Omega(\mathbf{r}; R, B) \rightarrow g(\mathbf{r}; B) \quad \text{when} \quad R \rightarrow 0 \qquad (6)$$

An important feature of function (5) is 'disorder transferability' which is similar to (3):

$$\Omega(\mathbf{r}; R, B) * g(\mathbf{r}; B_0) = \Omega(\mathbf{r}; R, B + B_0) \qquad (7)$$

In other words, knowing the decomposition (4) of an image of an immobile atom, $B_0 = 0$, we automatically get such decomposition for an image of an atom with any value $B_0$ of the displacement parameter.

It what follows, we note by $r = |\mathbf{r}|$ the distance and by $\bar{\Omega}(r; R, B)$ the radial component of the function $\Omega(\mathbf{r}; R, B)$. The decomposition procedure considers the radial component $\bar{f}(r)$ of the spherically symmetric function $f(\mathbf{r}) = \bar{f}(|\mathbf{r}|)$ being defined numerically, $f_n = \bar{f}(r_n)$, in the points $r_n$ of a regular grid $n = 0, 1, \ldots, N$; $r_0 = 0$; $r_n < r_{n+1}$, in the axis of real numbers. Generally speaking, the condition on the grid to be regular is not necessary but simplifies some calculations.

In the current version of the software, the discrepancy between the function $f(\mathbf{r})$ and its decomposition (4) is defined as the unweighted least-squares difference

$$LS(\{R_m, B_m, C_m\}) = \frac{1}{2} \sum_{n=0}^{N} [f_n - f_M(r_n; \{R_m, B_m, C_m\})]^2 \qquad (8)$$

where

$$f_M(r; \{R_m, B_m, C_m\}) = \sum_{m=1}^{M} C_m \bar{\Omega}(r; R_m, B_m) \qquad (9)$$

Other score functions can be easily implemented if required.

*1.2. The procedure*

The function (9) of a scalar variable $r$ includes the terms corresponding to the peaks, positive or negative. The number $M$ of terms is defined by the maximal authorized value, $M \leq M_{peaks}$, and by the accuracy $\varepsilon_{dec}$ of the decomposition which are parameters of the procedure. Various protocols have been tried mixing different order of the term assignment and refinement of estimated coefficients. Available software can use several such protocols, on the user choice. The default protocol consists in the following steps starting from the initial function values:

a) Find, in the increasing order of the argument $r$, all acceptable local peaks of the initial (at the first iteration) or the residual (at eventual following iterations) function; a peak $m$ in the position $r_n$ with $0 \leq n \leq N$, is selected if $|f_n| > \varepsilon_{dec}$ and $m \leq M_{peaks}$.
b) Estimate the coefficients of the respective terms from the shape of each peak.
c) Refine coefficients of all terms minimizing the score function (8).
d) Calculate the contribution (9) of the terms with the refined coefficients and extract this contribution from the initial function values.
e) If the break conditions defined above are not yet fulfilled, the procedure is repeated using the residual function; new terms complete the previous list and are refined, all together, at step c).

Alternative protocols break the procedure after the first iteration, whatever the residual discrepancy is, or refine the parameters values of each term before searching for the next one. The latter option may help to decompose the function with a very strongly dominating peak in the origin.

## 1.3. Peaks interpretation

The local peaks of the radial component are analyzed in the increasing order of the grid points $r_n$. When a peak with $|f_n| > \varepsilon_{dec}$, $0 \leq n \leq N$, is found, we first flip the function if this peak is negative. Then we identify the peak extension, $N_1 \leq n \leq N_2$ as discussed below

To estimate the initial values of the coefficients of the respective term $m$, we ignore the second term in (5) and a variation of $r$ around the peak. As a result, we impose $R_m = r_n$ and estimate two other coefficients from the minimum of

$$LS_{ln} = \sum_{n=N_1}^{N_2} \left[ ln(f_n) - ln\big(C_m \bar{g}(r_n; R_m, B_m)\big) \right]^2 \to \min_{R_m, B_m, C_m} \quad . \tag{10}$$

Here, for $R_m > 0$

$$\bar{g}(r; R_m, B_m) = \frac{1}{R_m^2} \sqrt{\frac{1}{4\pi B_m}} \exp\left( -\frac{4\pi^2 (r - R_m)^2}{B_m} \right) \tag{11}$$

and for $R_m = 0$

$$\bar{g}(r; 0, B_m) = \left(\frac{4\pi}{B_m}\right)^{3/2} \exp\left( -\frac{4\pi^2 r^2}{B_m} \right) \quad . \tag{12}$$

Comparison of logarithms instead of the function values allows one to get an analytic solution (Appendices A1 and A2).

If initially we flipped the function, we should invert the sign of the coefficient $C_m$.

## 1.4. Peak borders

The peak borders, $N_1 \leq n \leq N_2$, are chosen as the left and right function inflection points unless the function value drops earlier below $\varepsilon_{peak} > 0$ which is one more parameter of the procedure. Naturally, for the peak in the origin, $n = 0$, we assign $N_1 = 0$.

If a local maximum at the upper bound of the interval is observed, i.e., $|f(r_N)| > \varepsilon_{dec}$ and either $f(r_N) > f(r_{N-1}) > 0$ or $f(r_N) < f(r_{N-1}) < 0$, the actual peak of the curve $f(r)$ can be beyond the interval. By this reason, if the function values are known for $r_n > r_N$, they are used making this peak internal and reducing the situation to that described previously. By this reason, it is recommended to calculate the function, if possible, at an interval larger than that in which the decomposition is required.

If the function values beyond the interval are unknown, we assign $N_2 = R_m = r_N$.

If the peak is too narrow, $N_2 - N_1 \leq 1$, we assign $B_m, C_m$ such that the term

$C_m \bar{g}(r; R_m, B_m)$ in (8) becomes equal to $f(r_n)$ for $r = r_n$ and had its width equal to the grid interval.

## 1.5. Function calculation

Each term in (9) is a function with a single peak in the demi-axis $r \geq 0$, either an internal one or in the origin, when $8\pi^2 R^2 \leq 3B$ (**Fig. 1**). Its contribution is a decreasing function for $r > R$ and becomes negligibly small beyond some distance, i.e., $f_M(r; \{R_m, B_m, C_m\}) < \varepsilon_{term}$ with some small constant $\varepsilon_{term} \geq 0$ chosen. As soon as such point $r_n < r_N$ is found, the contribution of this term to the following points is ignored. A similar procedure is applied for $r < R$ moving from the peak toward the origin. If $\varepsilon_{term} = 0$, each term formally contributes to all points $r_n$ of the interval making the sum (9) more accurate but increasing computing time.

## 1.6. Refinement of parameter values

To optimize the fit (8) of the calculated function to the input one improving the initial estimates of the parameters of the decomposition (9), we use the standard minimizer L-BFGS (Lui & Nocedal, 1989). It requires calculation of the score function and of its gradient with respect to the decomposition coefficients (Appendix B).

The minimizer uses bottom bound restraints on the values of the coefficients, $R_{min} = 0$, $B_{min} > 0$ and $C_{min} > 0$. The values for the two latter bounds are calculated from the estimations for the minimal width and height of a peak giving a contribution equal to the precision threshold $\varepsilon_{dec}$.

For the peak in the origin, $R_m = 0$, we tried artificially to split it into two terms (5) with $R_m$ slightly different from zero (with estimated $C_m$ divided by two). In all examples we saw, $R_m$ always converged back to zero and we abandoned this practice.

## 1.7. Choice of control parameters

From our experience, when using a single term per Fourier ripple, the decomposition usually gives an approximation with the discrepancy of 3-4 orders of magnitude lower than the function values. As a consequence, refinement is sensitive to calculation errors (requiring double precision calculations in fortran) and to the initial estimations of the coefficients. While routinely the default choice of parameters gives sufficiently good results, in critical situations when a high accuracy, *i.e.*, a small $\varepsilon_{dec}$, is required they may be slightly improved after trying different $\varepsilon_{peak}$ and $\varepsilon_{term}$. The default values found empirically are $\varepsilon_{peak} = 5 \cdot 10^{-3}$ and $\varepsilon_{term} = 10^{-13}$; both values are defined as a part of the function value in the origin.

Except particular situations one does not use the maximal allowed number of terms $M_{peaks}$ ; the number of terms is found automatically for the given accuracy $\varepsilon_{dec}$. The last value can be defined either in the absolute scale or also as a part of the function value in the origin.

## 2. Program

*2.1. Software package, requirements and distribution*

The developed software consists of several programs. The principal component is the program *dec3D* which calculates the shell decomposition coefficients for a given oscillating function defined by its grid values. This program also calculates the sum of the respective terms and compares the calculated curve with the control one

An auxiliary program *dec3D-rsum* do the same but skips the first step and starts from a given set of the shell coefficients eventually refining them.

Since the principal goal is a decomposition of atomic images, the software includes one more auxiliary program *dec3D-atom* that calculates the atomic images for the selected types of atoms for the chosen values of the resolution and the isotropic displacement factor.

The programs are available in fortran77 and in python3 (version 3.7.4 was used to develop and test the programs). The fortran version uses the L-BFSG subroutines provided with the program; thus, the program is independent of the environment. The fortran version has no specific requirements. For example, working under Windows, one can use the compiler *g77* with no installation needed.

The python version also uses only basic constructions. It requires libraries *numpy* (version 1.18.4 or higher) and *scipy* (version 1.6.3 or higher) while may be compatible with some older versions. The L-BFGS minimizer is distributed with the *scipy* libraries; as a consequence, the results may slightly vary with the installed versions of the libraries.

The python scripts may be run as stand-alone of via a graphical interface. The GUI python version required additionally *wx-python* (version 4.0.6 or higher) and *MatPlotLib* (version 3.4.3 or higher).

Some details of the programs are discussed below. The programs can be obtained by request from the authors or from the site

https://ibmc.cnrs.fr/en/laboratoire/arn-en/presentation/structures-software-and-websites/

*2.2. Main program*

*2.2.1. Input data*

The principal program *dec3D* requires the function values which are defined in the input file containing the records with the distance value and the function values in these points. The file may contain any number of columns with the values of different functions. For example, these may be images of atoms of different types at a given resolution. Otherwise, they may be images of atoms with different values of the displacement parameter or at a different resolution, or something else.

The program requires also a file of input parameters where the name of the file with the function values (the file discussed above) is the only mandatory parameter; optional parameters define the file content and the choice of the appropriate columns with data. Other optional parameters define the maximal allowed number of terms, the maximal allowed discrepancy between the function and its decomposition, and the parameters used to build the terms and to find the range of their contribution, as discussed in previous sections.

*2.2.2. Output data*

The principal output file contains the initial estimations of $R_m, B_m, C_m$ of the shell decomposition (4) and their values after refinement.

The program calculates the function (8) as the sum of the terms as they are reported at the end of the program. Also, the difference function between the initial function and its decomposition (9) is calculated for all grid points in which the function is defined. These functions are saved in a separated output file and may be plotted and analyzed. The maximal deviation value is communicated for the whole interval of the function definition and for the interval for which the decomposition is constructed.

*2.3. Auxiliary programs*

*2.3.1. Calculation of atomic images*

The auxiliary program *dec3D-atom* calculates the images of scatterers of a selected type for any chosen resolution and any value of the isotropic displacement parameter, equal to zero by default. These images are calculated as the Fourier transform of the respective scattering functions (Appendix C).

X-ray diffraction or electron scattering functions of neutral atoms and their ions are defined by tabulated functions of $s = d^{-1} = 2 \sin \theta / \lambda$ (*e.g.*, Brown *et al.*, 2006) or by a multi-Gaussian approximation to them (Navaza, 1994; Peng, 1999; Brown *et al.*, 2006, and references therein). The choice of the appropriate file is defined in the input parameters. Any other data in the same format, *e.g.*, Waasmaier & Kirfel (1995) or Grosse-Kunstleve *et al.* (2004), may be used if available. If necessary, the files may be completed by scattering functions for any new kind of scattering factor, including that for a single Gaussian used at subatomic resolution (Afonine *et al.*, 2007).

*2.3.2. Calculation of the resulted curves*

The auxiliary program *dec3D-rsum* reads the values of the coefficients of the shell decomposition and calculates the sum (9) as well as the difference between it and the input function. Differently from the principal program, this one starts directly from a set of the decomposition coefficients allowing one to test their different combinations. Optionally, the input values of the coefficients can be refined minimizing (8).

A program *dec3D-sum* is a simplified version of *dec3D-rsum* which calculates the sums (8) with the given values of the decomposition coefficients without their refinement. Differently from *dec3D-rsum*, this program needs neither *scipy* library for the python version, nor the L-BFGS modules for the Fortran version.

*2.4. GUI program version*

The computational part of the *dec3D* python suite is incorporated into an interactive graphic interface. This GUI version ignores some advanced options of the stand-alone programs.

Instead, it gives some extra possibilities such as a calculation 'on fly' of an analytic function to be decomposed.

The GUI program runs consecutively four components:

a) Get curves to be decomposed;
b) Get the decomposition coefficients;
c) Calculate the decomposition curves with the decomposition coefficients which can be eventually modified;
d) Plot any combination of the calculated curves and optionally save them.

One can save intermediate results and to come back to any of the previous steps.

At the first step, three options are available. One can input some precalculated functions or calculate atomic images of a given resolution and with a given displacement parameter. The atomic types, or those of respective ions, are selected from the interactive Mendeleev's Table. Finally, one can define the input function by a python line in the respective window as, for example,

3*(math.sin(2*math.pi*x) - (2*math.pi*x)*math.cos(2*math.pi*x)) / ((2*math.pi*x)**3)

for the radial component of the three-dimensional interference function

$$G(\mathbf{x}) = 3 \frac{\sin(2\pi|\mathbf{x}|) - (2\pi|\mathbf{x}|)\cos(2\pi|\mathbf{x}|)}{(2\pi|\mathbf{x}|)^3} \qquad (13)$$

At the second step, the principal program works decomposing the functions obtained at the previous step.

The third step allows one to modify the values of the coefficients, exclude some terms from the curve calculation or, inversely, add manually new terms. Then, the sum (9) is calculated.

At the fourth step, any combination of the calculated functions and their difference can be plotted, analyzed visually and saved in a file. One can choose the curve attributes such as their type, width and color.

**Fig. 2** illustrates, step by step, calculation of the images of the carbon and sulphur atoms, with $B = 0$ and at a resolution of 3 Å, using the GUI version of the program. The plots are generated directly by the program.

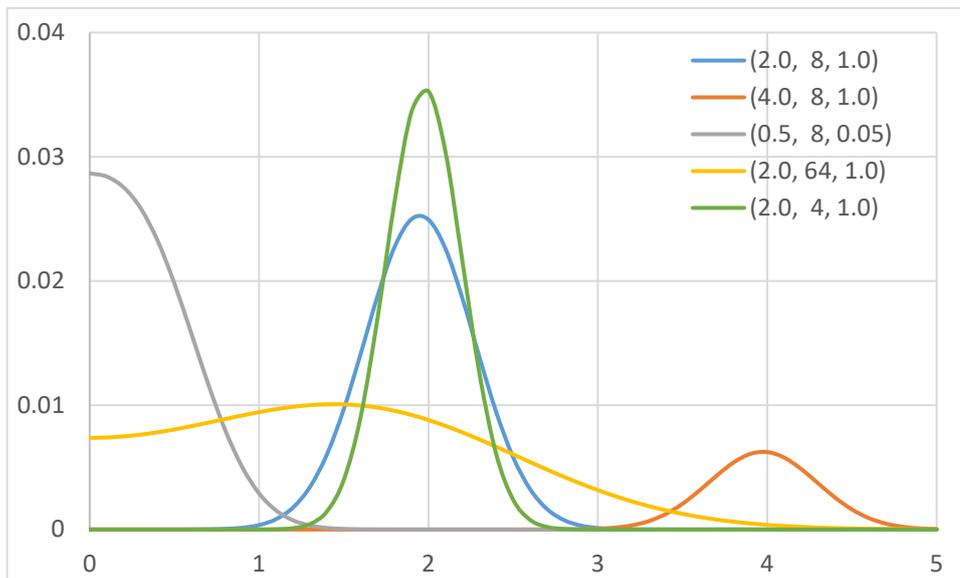

**Figure 1.** The radial part $C\bar{\Omega}(r; R, B)$ of the scaled shell function for several sets of parameters $(R, B, C)$ as indicated in the legend.

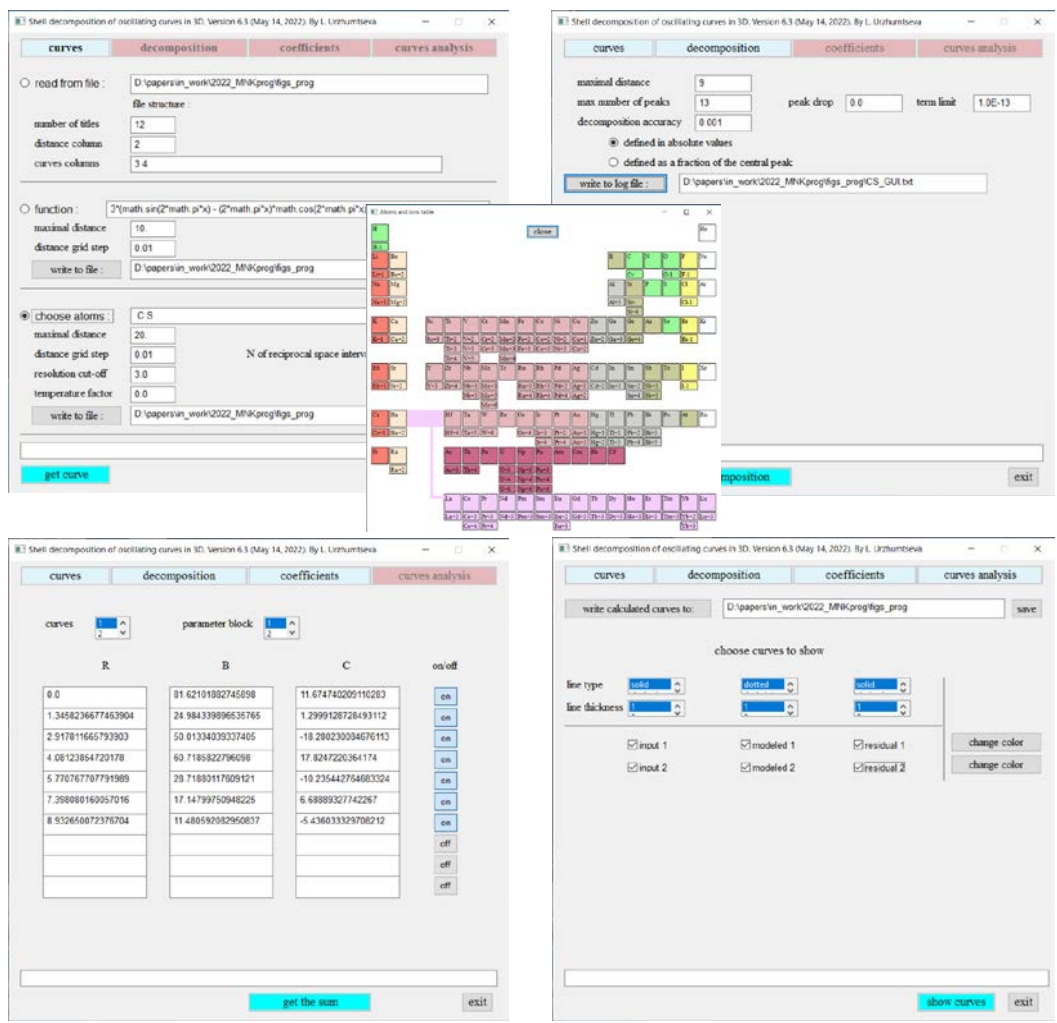
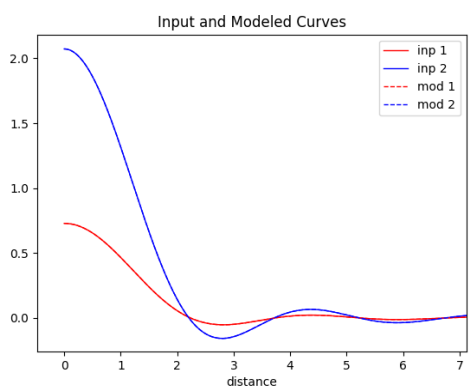
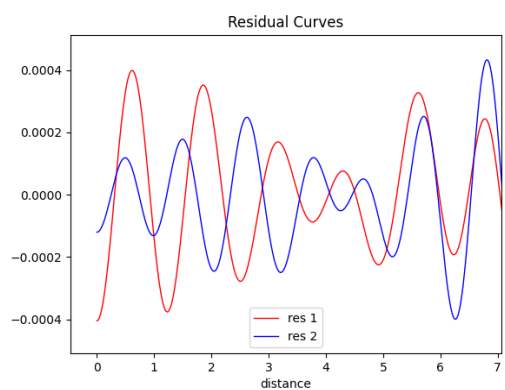

**Fig. 2.** Illustration of calculating the atomic images using the GUI version of the software. Consecutive windows show the selection of atomic types and parameters of the images, the superposed interactive Mendeleev table, calculated decomposition parameters, windows for plotting the curves, the curves themselves and the residual curved after decomposition (note the difference in scale in two last figures). Blue curves correspond to the images of carbon, red curves to those for sulphur. In the left bottom image, broken lines for the decomposition are indistinguishable from the full curves for the input images.

**Acknowledgement**

AU acknowledges Instruct-ERIC and the French Infrastructure for Integrated Structural Biology FRISBI [ANR-10-INBS-05].

**References**

Afonine, P.V., Grosse-Kunstleve, R.W., Adams, P.D., Lunin, V.Y. & Urzhumtsev, A.G. (2007). *Acta Cryst.,* D**63**, 1194-1197.

Afonine, P.V., Poon, B.K., Read, R.J., Sobolev, O.V., Terwilliger, T.C., Urzhumtsev, A.G. & Adams, P.D. (2018). *Acta Cryst.,* D**74**, 531-544.

Atkinson, K.E. (1989). *An Introduction to Numerical Analysis (2nd ed.). John Wiley & Sons.* ISBN 0-471-50023-2.

Baek, M. *et al.* (2021). *Science*, **373**, 871–876.

Brown, P.J., Fox, A.G., Maslen, E.N., O'Keefe, M.A. & Willis, B.T.M. (2006). In: Prince E, editor. *International Tables for X-ray Crystallography*, Vol. C, Dordrecht: Springer, Dordrecht, The Netherlands. ch. 6.1, pp. 554-595.

Chapman, M.S. (1995). *Acta Cryst.* A**51**, 69-80.

Chapman, M. S., Trzynka, A. & Chapman, B. K. (2013). *J. Str. Biol.* **182**, 10-21.

Diamond, R. (1971). *Acta Cryst*. A**27**, 436-452.

Doyle, P. A. & Turner, P. S. (1968). *Acta Cryst*. **24**, 390-397.

Grosse-Kunstleve, R.W., Sauter, N.K. & Adams, P.D. (2004). *Newsletter of the IUCr Commission on Crystallographic Computing*, 3, 22-31.

Jumper, J. *et al.* (2021). *Nature* **596**, 583-589.

Kühlbrandt, W. (2014). *Science* **343**, 1443-1444.

Liu, D. C. & Nocedal, J. (1989). *Math. Program*. **45**, 503–528.

Navaza, J. (1994). *Acta Cryst*. A**50**, 157-163.

Palmer, C.M. & Aylett, C.H.S. (2022). *Acta Cryst.* D**78,** 136-143.

Peng, L.-M. (1999). *Micron,* **30**, 625-648.

Urzhumtsev, A.G. & Lunin, V.Y. (2019). *Crystallogr. Reviews* **25**, 164-262.

Urzhumtsev, A.G. & Lunin, V.Y. (2022). *BioRxiv*, 10.1101/2022.03.28.486044

Waasmaier, D. & Kirfel, A. (1995). *Acta Cryst*. A**51**, 416-431.

# Appendix A. Parameters of the decomposition terms

## A1. Gaussian approximation to a local peak

To find the best Gaussian approximation $U \exp(-vr^2)$, with $U, v$ constants, to the given set of values $f_n = f(r_n)$, $N_1 \leq n \leq N_2$, we introduce first intermediate parameters $y_n = ln(f_n)$, $u = lnU$, with which condition (10), with the parameters introduced as below in A2, becomes

$$LS_2(u, v) = \sum_{n=N_1}^{N_2} [(u - vr_n^2) - y_n]^2 \to \min_{u,v} \tag{A1}$$

i.e.,

$$\frac{\partial y}{\partial u} LS_2(u, v) = \frac{\partial y}{\partial v} LS_2(u, v) = 0 \tag{A2}$$

This gives a system of two linear equations with respect to $u, v$ resulting in

$$\begin{cases} u = (sr2 \cdot syr2 - sr4 \cdot sy)(sr2 \cdot sr2 - n12 \cdot sr4)^{-1} \\ v = (n12 \cdot syr2 - sr2 \cdot sy)(sr2 \cdot sr2 - n12 \cdot sr4)^{-1} \end{cases} \tag{A3}$$

with $n12 = N_2 - N_1 + 1$ and

$$sr2 = \sum_{n=N_1}^{N_2} r_n^2; \quad sr4 = \sum_{n=N_1}^{N_2} r_n^4; \quad sy = \sum_{n=N_1}^{N_2} y_n; \quad syr2 = \sum_{n=N_1}^{N_2} y_n r_n^2 \tag{A4}$$

## A2. Estimation of the decomposition parameters

For the peak in the origin, the optimal values of the constants $u, v$ in (A1) are related to the $B, C$ parameters as

$$u = lnC + \frac{3}{2}ln(4\pi) - \frac{3}{2}ln(B) \; ; \; v = 4\pi^2 B^{-1} \tag{A5}$$

giving

$$B = 4\pi^2 v^{-1} \; ; \; C = \pi^{3/2} v^{-3/2} e^u \tag{A6}$$

To calculate $B, C$ from $u, v$ for an internal peak, we consider $y_n = ln(2f_n R^2 \pi^{1/2})$ ignoring the variation of the parameter $r$ in the interval $r_{N_1} \leq n \leq r_{N_2}$. Then, the optimal values of the constants $u, v$ in (A1) are related to the $B, C$ parameters as

$$u = lnC - \frac{1}{2}ln(B) \; ; \; v = 4\pi^2 B^{-1} \tag{A7}$$

giving

$$B = 4\pi^2 v^{-1}; C = 2\pi u v^{-1/2} \quad (A8)$$

If a peak is too narrow, with its region composed of only 1 or 2 points, we assign $B = B_{min}$, the minimal allowed value. Then, from the condition $C\bar{g}(0; R = 0, B) = f_{residual}(r_0)$ for the peak in the origin, we calculate $C = f(r_0)B^{3/2}\pi^{-3/2}$, and, from the condition $C\bar{g}(r_{N0}; R, B) = f_{N0}$ for an internal peak, we calculate $C = 2f_{N0}R^2(\pi B)^{1/2}$.

## Appendix B. $\Omega$-function and its derivatives

### B1. Approximation to the function near the origin

The contribution of a term $\bar{\Omega}(r; R, B)$ with $R \neq 0$ to the curve in the point $r = 0$ is calculated according to the asymptotic behavior of function near the origin:

$$\bar{\Omega}(r; R, B) = \frac{1}{rR}\sqrt{\frac{1}{4\pi B}} \exp\left(-\frac{4\pi^2(r^2 + R^2)}{B}\right)\left[\exp\left(\frac{8\pi^2 rR}{B}\right) - \exp\left(-\frac{8\pi^2 rR}{B}\right)\right] =$$

$$= \frac{1}{rR}\sqrt{\frac{1}{4\pi B}} \exp\left(-\frac{4\pi^2(r^2 + R^2)}{B}\right)\left[\frac{16\pi^2 rR}{B} + O(r^3)\right]$$

$$= \left(\frac{4\pi}{B}\right)^{3/2} \exp\left(-\frac{4\pi^2 R^2}{B}\right)[1 + O(r^2)] \quad (B1)$$

### B2. Partial derivatives of the score function

Partial derivatives of the score function $D(\{R_m, B_m, C_m\})$ describing the fit of the decomposition to the input curve are required by the minimizer L-BFSG (Liu & Nocedal, 1989). These derivatives are calculated using the chain rule, in particular for (8):

$$\frac{\partial}{\partial p} LS(\{R_m, B_m, C_m\}) = \sum_{n=0}^{N} [f_n - f_M(r_n; \{R_m, B_m, C_m\})]\frac{\partial f_M(r_n; \{R_m, B_m, C_m\})}{\partial p} \quad (B2)$$

where $p$ is any of the coefficients $R_m, B_m, C_m$.

A straightforward differentiation of the expressions for the general case $R_m \neq 0, r \neq 0$

gives

$$\frac{\partial f_M(r;\{R_m,B_m,C_m\})}{\partial R_m} = \frac{C_m B_m^{-3/2}}{2\pi^{1/2} r R_m^2} \times \quad (B3)$$

$$\left\{\left[\exp\left(-\frac{4\pi^2(r-R_m)^2}{B_m}\right)[8\pi^2 R_m(r-R_m)-B_m]\right.\right.$$

$$\left.\left. + \exp\left(-\frac{4\pi^2(r+R_m)^2}{B_m}\right)\right][8\pi^2 R_m(r+R_m)+B_m]\right\}$$

$$\frac{\partial f_M(r;\{R_m,B_m,C_m\})}{\partial B_m} = \frac{C_m B_m^{-5/2}}{4r R_m \pi^{1/2}} \times \quad (B4)$$

$$\left\{\exp\left(-\frac{4\pi^2(r-R_m)^2}{B_m}\right)[8\pi^2(r-R_m)^2-B_m]\right.$$

$$\left. - \exp\left(-\frac{4\pi^2(r+R_m)^2}{B_m}\right)[8\pi^2(r+R_m)^2-B_m]\right\}$$

$$\frac{\partial f_M(r;\{R_m,B_m,C_m\})}{\partial C_m} = \quad (B5)$$

$$\frac{1}{rR_m}\sqrt{\frac{1}{4\pi B_m}}\left[\exp\left(-\frac{4\pi^2(r-R_m)^2}{B_m}\right) - \exp\left(-\frac{4\pi^2(r+R_m)^2}{B_m}\right)\right]$$

For $R_m = 0, r \neq 0$, *i.e.*, for the Gaussian terms, these derivatives are:

$$\frac{\partial f_M(r;\{0,B_m,C_m\})}{\partial B_m} = 4C_m \pi^{3/2} B_m^{-7/2} \exp\left(-\frac{4\pi^2 r^2}{B_m}\right)[8\pi^2 r^2 - 3B_m] \quad (B6)$$

$$\frac{\partial f_M(r;\{0,B_m,C_m\})}{\partial C_m} = \left(\frac{4\pi}{B_m}\right)^{3/2} \exp\left(-\frac{4\pi^2 r^2}{B_m}\right) \quad (B7)$$

Finally, for $R_m \neq 0, r = 0$ the derivatives are:

$$\frac{\partial f_M(0;\{R_m,B_m,C_m\})}{\partial R_m} = -64\pi^{\frac{7}{2}} C_m R_m B_m^{-\frac{5}{2}} \exp\left(-\frac{4\pi^2 R_m^2}{B_m}\right) \quad (B8)$$

$$\frac{\partial f_M(0;\{R_m,B_m,C_m\})}{\partial B_m} = 4\pi^{3/2} C_m B_m^{-7/2}(8\pi^2 R_m^2 - 3B_m)\exp\left(-\frac{4\pi^2 R_m^2}{B_m}\right) \quad (B9)$$

$$\frac{\partial f_M(0;\{R_m,B_m,C_m\})}{\partial C_m} = \left(\frac{4\pi}{B_m}\right)^{3/2} \exp\left(-\frac{4\pi^2 R_m^2}{B_m}\right) \quad (B10)$$

# Appendix C. Calculations of the atomic images

The image of an atom with an isotropic scattering factor $f(s)$ and calculated with the resolution cut-off $D = d_{high}$

$$\rho_n(\mathbf{r}; B, D) = \int_{s \leq D^{-1}} f_n(s) \exp(-Bs^2/4) \exp(-2\pi i \mathbf{rs}) \, d\mathbf{s}$$

is a spherically symmetric function. Here $\mathbf{rs}$ is a dot product of two respective vectors. We calculate the radial component of this function along the axis Oz using spherical coordinates which gives

$$\bar{\rho}_n(r; B, D) = 2r^{-1} \int_0^{D^{-1}} s f_n(s) \exp(-Bs^2/4) \sin(2\pi rs) \, ds$$

When $r \ll 1$,

$$\bar{\rho}_n(r; B, D) \approx 4\pi \int_0^{D^{-1}} s^2 f_n(s) \exp(-Bs^2/4) \, ds$$

The Simpson (also known as Kepler-Simpson or Newton-Cotes) quadrature (*e.g.*, Atkinson, 1989) allows a numerical calculation of these integrals in any fine grid.